\documentclass{svproc}
\usepackage{url}

\usepackage{graphicx}
\usepackage{wrapfig}
\usepackage{caption}
\usepackage{hyperref}

\begin{document}
\mainmatter              
\title{Some studies using capillary for flow control in a closed loop gas recirculation system}
\titlerunning{Capillary Study}  
%
\author{S.D.Kalmani\inst{1} \and S. Mondal\inst{1}
R.R. Shinde\inst{1} \and P.V.Hunagund\inst{2}}
\authorrunning{S.D.Kalmani et al.} 
%
%
\institute{Tata Institute of Fundamental Research, Mumbai, India,\\
\email{kalmani@tifr.res.in},\\ 
\and
Department of Applied Electronics and Research,
Gulbarga University, Gulbarga, India\\}

\maketitle              

\begin{abstract}
A Pilot unit of a closed loop gas (CLS) mixing and distribution system for the INO project \cite{inoReport} was designed and is being operated with $(1.8 \times 1.9)$ $m^2$ glass RPCs (Resistive Plate Chamber). The performance of an RPC \cite{gas:recirculate} depends on the quality and quantity of gas mixture being used, a number of studies on controlling the flow and optimization of the gas mixture is being carried out. 
In this paper the effect of capillary as a dynamic impedance element on the differential pressure across RPC detector in a closed loop gas system is being highlighted. The flow versus the pressure variation with different types of capillaries and also with different types of gasses that are being used in an RPC is presented. An attempt is also made to measure the transient time of the gas flow through the capillary.
\end{abstract}
\section{Introduction}
\begin{figure}
	\centering
	\begin{minipage}{0.5\textwidth}
		\centering
		\includegraphics[width=0.95\linewidth]{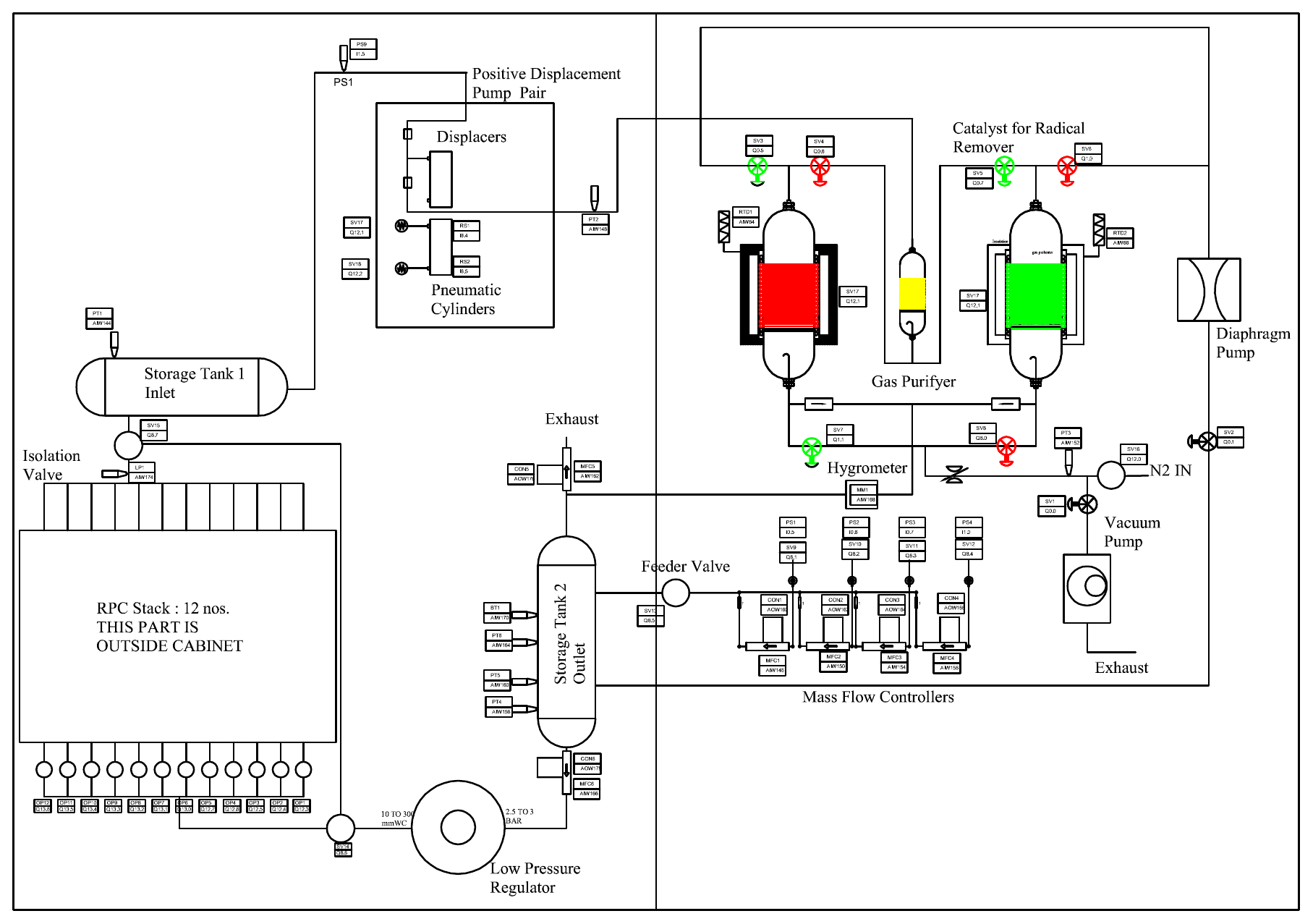}
		\captionsetup{width=0.8\linewidth}
		\captionof{figure}{Block Diagram of CLS}
		\label{fig:CLS}
	\end{minipage}%
	\begin{minipage}{.5\textwidth}
		\centering
		\includegraphics[width=0.95\linewidth]{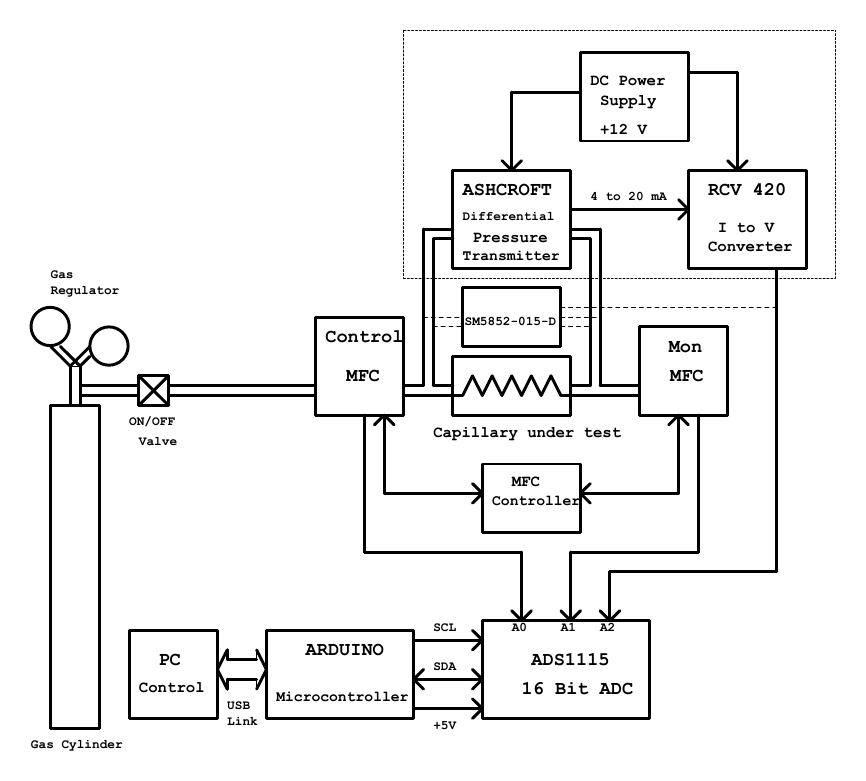}
		\captionsetup{width=0.8\linewidth}
		\captionof{figure}{The Experimental Setup}
		\label{fig:ExperimentlaSetup}
	\end{minipage}
\end{figure}

In the initial design of the CLS, as shown in figure \ref{fig:CLS}, a high pressure to low pressure [HPLP] diaphragm based device was used to regulate, the high pressure (100 Kpa) from the storage tank to low pressure (200 pa) feeding to the RPCs. Due to continuous atmospheric pressure variation cycle and low flow rate of a few SCCM (Standard Cubic Centimeter per Minute) in glass RPCs, it was necessary for the pressure controller to adjust pressure so as not to exceed the safe operating differential pressure. However the HPLP regulator was found not responding as quickly as required to compensate pressure changes. This was mainly due to frictional and inertial hysteresis, which is common to mechanical dynamic devices. Therefore, a flow limiter namely a simple capillary in lieu of the regulator was used at input of the each RPC and also an external pressure sensor in the room to correct the pressure inside the CLS was implemented. But to ensure a uniform gas flow through the RPC and to connect several RPCs in series, a detailed study on the use of the capillary has been carried out.


The experimental setup, as shown in figure \ref{fig:ExperimentlaSetup}, consists of two MFCs (Mass Flow Controllers) and differential pressure sensors namely ASHCOFT XLdp and a Silicon Microstructures Inc. (SM5852). A microcontroller ARDUINO board with an additional four channel ADC module is designed to read data of the pressure difference across the capillary at a rate of 60 Hz. Two MFCs, one for controlling the flow and the other for reading the flow are used and these are calibrated using water displacement method for each of the gases. The full scale range of the MFCs is 45 SCCM which is in the range of interest.

\paragraph{Types of capillaries fabricated:}
Using Poiseuille's law for fluids and assuming the input pressure to be 1000 mbar, the length of stainless (SS) tube to be 2500 mm, pressure difference required to be say about 2.5 mbar for a flow rate of 6 SCCM and considering the viscosity of the R134a gas a major component, the calculated diameter of the capillary tube is 300 microns. This tube is wound on a bobbin with a quarter inch tube on either side.

The studies are performed using gasses that are used in the glass RPCs namely $C_2H_2F_4$ (R 134a), $iC_4H_{10}$ (Isobutane) and $SF_6$ (Sulphur hexafluoride).

\section{Results and Conclusions}
\begin{figure}
	\centering
	\begin{minipage}{0.5\textwidth}
		\centering
		\includegraphics[width=0.95\linewidth]{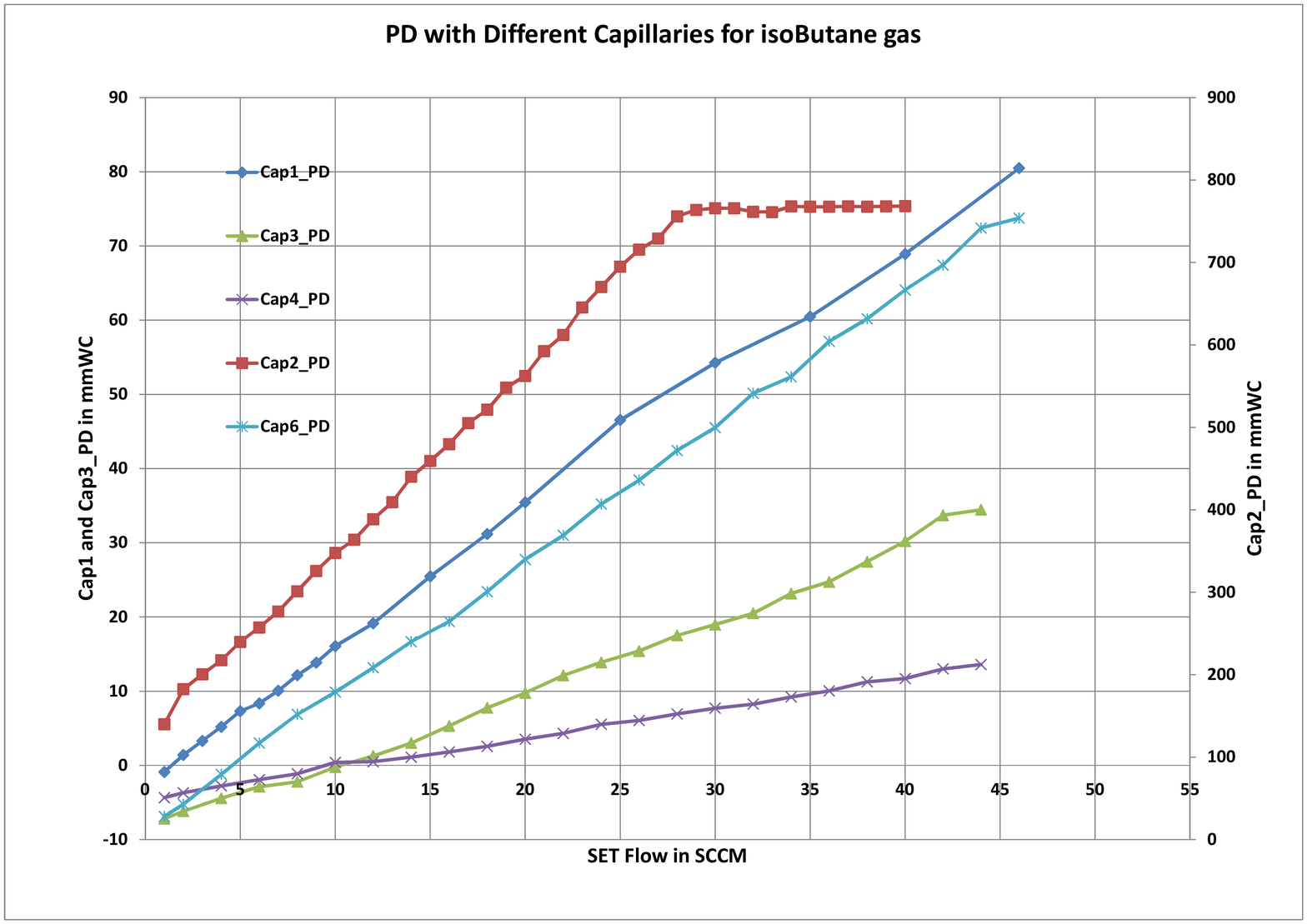}
		\captionsetup{width=0.85\linewidth}
		\captionof{figure}{Linear flow in the region of interest (Ashcroft Sensor)}
		\label{fig:AllCaps}
	\end{minipage}%
	\begin{minipage}{.5\textwidth}
		\centering
		\includegraphics[width=0.95\linewidth]{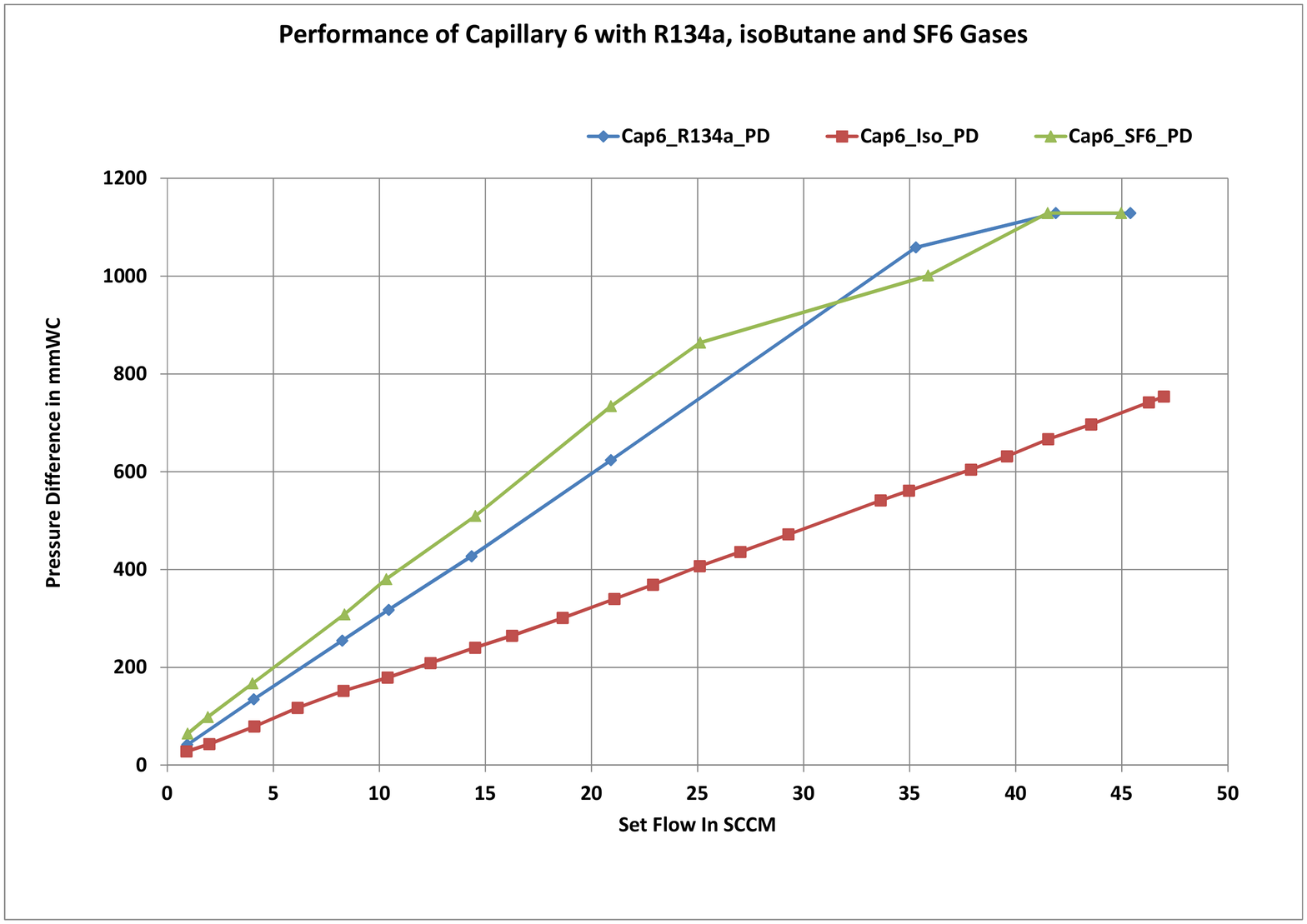}
		\captionsetup{width=0.85\linewidth}
		\captionof{figure}{Linear flow up to 25 SCCM with different gases of interest}
		\label{fig:C6AllGases}
	\end{minipage}
\end{figure}

It has been observed that capillaries with different diameter and length offer different impedance to gas flow. If we consider the flow for Isobutane gas then, as seen in figure \ref{fig:AllCaps}, a linear flow up to 28 SCCM is seen for capillaries C1, C2, C3 and C4. The flow resistance depends on the type of gas that flows through it, higher the density gas, higher is pressure required to obtain a given flow rate, as seen in figure \ref{fig:C6AllGases}. Here isobutene has lowest density among the three gases, hence it produces lowest pressure drop for a given flow. The capillaries C1 and C3 capillary do not qualify for glass RPC applications where the flow is of the order of a few SCCM. The capillary C2 and C6 show ideal performance and could be most suitable one for INO RPCs. The transient time measured by the existing system for R134a gas is about 335 microseconds, but for precise measurement we need to have a different tool.

%
%


\begin{thebibliography}{2}
%

\bibitem {inoReport}
INO-Collaboration, India-based neutrino observatory, Tech. Rep.
INO/2006/01, Tata Institute of Fundamental Research (January 2006).
\url{http://www.ino.tifr.res.in/ino/OpenReports/INOReport.pdf}

\bibitem {gas:recirculate}
M. Bhuyan, et. al., Performance of the prototype gas recirculation
system with built-in RGA for INO RPC system. In: Nuclear Instruments
and Methods in Physics Research Section A: Accelerators, Spectrometers,
Detectors and Associated Equipment 661, Supplement 1 (2012) S234 
170 S240, x. Workshop on Resistive Plate Chambers and Related Detectors
(RPC 2010). \url{doi:http://dx.doi.org/10.1016/j.nima.2010.09.169}

\end{thebibliography}
\end{document}